\documentclass[twocolumn, superscriptaddress, 10pt, nofootinbib, notitlepage]{revtex4-2}
\usepackage[utf8]{inputenc}
\usepackage{amsmath}
\usepackage{amsfonts}
\usepackage{amssymb}
\usepackage{braket}
\usepackage{physics}
\usepackage{dsfont}
\usepackage{graphicx}
\usepackage{float}
\usepackage{array}
\usepackage[normalem]{ulem}
\usepackage{tikz}
\usetikzlibrary{positioning,calc,fit,shapes,backgrounds,graphs,matrix,arrows,arrows.meta,decorations.pathreplacing,decorations.pathmorphing}
\usepackage[caption=false]{subfig}
\usepackage{natbib}
\begin{document}
\title{
Entanglement Thresholds of Doubly-Parametric Quantum Transducers
}

\author{Curtis L. Rau}
\thanks{These authors contributed equally}
\affiliation{Department of Physics, University of Colorado, Boulder, Colorado 80309, USA}
\affiliation{National Institute of Standards and Technology (NIST), Boulder, Colorado 80305, USA}
\author{Akira Kyle}
\thanks{These authors contributed equally}
\affiliation{Department of Physics, University of Colorado, Boulder, Colorado 80309, USA}
\affiliation{National Institute of Standards and Technology (NIST), Boulder, Colorado 80305, USA}
\author{Alex Kwiatkowski}
\affiliation{Department of Physics, University of Colorado, Boulder, Colorado 80309, USA}
\affiliation{National Institute of Standards and Technology (NIST), Boulder, Colorado 80305, USA}
\author{Ezad Shojaee}
\affiliation{National Institute of Standards and Technology (NIST), Boulder, Colorado 80305, USA}
\author{John D. Teufel}
\affiliation{National Institute of Standards and Technology (NIST), Boulder, Colorado 80305, USA}
\author{Konrad W. Lehnert}
\affiliation{Department of Physics, University of Colorado, Boulder, Colorado 80309, USA}
\affiliation{National Institute of Standards and Technology (NIST), Boulder, Colorado 80305, USA}
\affiliation{JILA, University of Colorado and NIST, Boulder, Colorado 80309, USA}
\author{Tasshi Dennis}
\affiliation{National Institute of Standards and Technology (NIST), Boulder, Colorado 80305, USA}
\date{\today}

\begin{abstract}
Doubly-parametric quantum transducers, such as electro-opto-mechanical devices, are quickly approaching quantum operation as decoherence mechanisms such as thermal noise, loss, and limited cooperativities are improved.
These devices show potential as the critical link between quantum information contained at frequencies as disparate as those in the optical and microwave domains, thus enabling applications such as long distance networking of superconducting quantum computers.
However, the requirements on the operating parameters of the transducers necessary to achieve quantum operation have yet to be characterized.
In this work we find simple, explicit expressions for the necessary and sufficient conditions under which doubly-parametric transducers in the resolved-sideband, steady-state limit are capable of entangling optical and microwave modes.
Our analysis treats the transducer as a two-mode bosonic Gaussian channel capable of both beamsplitter-type and two-mode squeezing-type interactions between optical and microwave modes.
For the beamsplitter-type interaction, we find parameter thresholds which distinguish regions of the channel's separability, capacity for bound entanglement, and capacity for distillable entanglement.
By contrast, the two-mode squeezing-type interaction always produces distillable entanglement with no restrictions on temperature, cooperativities, or losses.
Finally, we find the entanglement breaking conditions on the reduced one-mode upconversion and downconversion channels obtained by initializing the unused input with vacuum while ignoring the unused output.
These differences between the entanglement thresholds of the beamsplitter-type and two-mode squeezing-type interactions are then important considerations in the construction of larger quantum networks that integrate multiple transducers.
\end{abstract}

\maketitle
\section{Introduction}

The realization of quantum transduction between the microwave frequencies of solid state qubits and the optical wavelengths of current low-loss communications is a critical step towards building large-scale, long-distance quantum networks \cite{lauk20_persp_quant_trans}.
Coherently bridging the six orders of magnitude separating these frequencies with minimal loss and added noise remains a technical challenge for all the various transducer implementations that have been proposed and are being actively pursued \cite{andrews14_bidir_effic_conver_between_microw_optic_light, jiang20_effic_bidir_piezo_optom_trans, rueda19_elect_optic_entan_sourc_microw, fan18_super_cavit_elect_optic, williamson14_magnet_optic_modul_with_unit_quant_effic, safavi-naeini11_propos_optom_travel_wave_phonon_photon_trans, chu20_persp_hybrid_quant_opto_elect_system}.
Therefore, given the parameters describing the performance of an optical--electrical (OE) transducer (such as cooperativities, noises, and losses), the task is to determine what thresholds need to be exceeded in order for the transducer to successfully operate within a quantum network.

OE transduction will likely significantly limit the performance of near-term quantum networks.
So understanding when a transducer is not capable of supporting any possible quantum communication protocol is a crucial step in designing and evaluating protocols which are robust to transduction imperfections.
We define the successful quantum operation of a transducer to be its ability to connect the optical and microwave domains through entanglement, which is only possible when the quantum channel describing a transducer is not separable across the OE partition.
This is because a transducer whose channel is separable will never output entangled states and so any state it outputs can always be prepared by local operations separately in the optical and microwave domains with classical communication between them.
Thus, this definition constitutes a necessary, but not always sufficient, condition that any OE transducer must first satisfy before it is capable of facilitating quantum communication between the optical and microwave domains.
As we will show, there are some operating regions where a transducer is only capable of establishing bound (non-distillable) OE entanglement, which describes entanglement where many copies of the state cannot be reduced to a single, more highly entangled and purer state using local operation and classical communication (LOCC) alone \cite{horodecki09_quant_entan}.
Therefore, it is unlikely that a quantum communication protocol would be able to achieve high fidelity quantum communication through such a transducer, even though under this definition, it is operating in the quantum regime.

Previous work has analyzed and given explicit protocols for faithful conversion of OE modes using various types of transducers.
In the simplest protocol, the mode to be converted is sent through the port in one domain and ideally output from the port in the other domain.
This scheme implicitly initializes the unused input port with vacuum while the unused output is ignored (traced over).
The resulting quantum channel can always be characterized by effective loss and added noise parameters which directly determine the transduction fidelity \cite{zeuthen20_figur_merit_quant_trans}.
An alternative protocol uses the OE entanglement generation capability of the transducer itself as a resource to accomplish the mode conversion via quantum teleportation \cite{barzanjeh11_entan_optic_microw_cavit_modes, barzanjeh12_rever_optic_to_microw_quant_inter, zhong20_propos_heral_gener_detec_entan, zhong20_entan_microw_optic_modes_stron}.
More involved protocols introduce squeezed ancilla states to the unused input ports and homodyne measurements of the unused output ports in order to improve transduction fidelity of imperfect transducers \cite{zhang18_quant_trans_with_adapt_contr}.
Furthermore, allowing multiple uses of the transduction channel interspersed with single mode unitaries can also improve transduction fidelity of imperfect transducers with reduced need for squeezing and homodyne measurement resources \cite{lau19_high_fidel_boson_quant_state, zhang20_inter_based_univer_decoup_swapp}.
In all of these protocols, the classical average fidelity bounds for the transduced states cannot be surpassed if the transducer fails to achieve quantum operation under our definition.
Thus finding the thresholds for when a transducer achieves quantum operation will set the lower bounds for when any of these protocols can succeed or when protocols for networking remote microwave modes over optical links become limited by the transducers at each node \cite{stannigel10_optom_trans_long_distan_quant_commun, stannigel11_optom_trans_quant_infor_proces, krastanov20_optic_heral_entan_super_system_quant_networ, abdi14_entan_two_distan_non_inter_microw_modes, cernotik16_measur_induc_long_distan_entan, yin15_quant_networ_super_qubit_throug_optom_inter}.

Broadly, OE transducers can be classified as either singly-parametric or doubly-parametric transducers.
The class of singly-parametric transducers utilize a single three-wave mixing interaction with one optical pump to facilitate the information exchange and can be physically implemented in Pockels effect materials \cite{rueda19_elect_optic_entan_sourc_microw, fan18_super_cavit_elect_optic}, atomic ensembles \cite{williamson14_magnet_optic_modul_with_unit_quant_effic}, magnons in ferromagnetic crystals \cite{hisatomi16_bidir_conver_between_microw_light}, quantum dots coupled to surface acoustic waves \cite{autry20_surfa_acous_wave_cavit_inas, weiss21_optom_wave_mixin_by_singl_quant_dot}, and piezo-optomechanical crystals \cite{jiang20_effic_bidir_piezo_optom_trans}. 
Whereas doubly-parametric transducers (DPTs) utilize a bosonic mediating mode which is coupled to both the optical and microwave modes through two three-wave mixing interactions and require both an optical and microwave pump -- effectively implementing a four-wave mixing interaction between the two pumps and two signal modes.
Electro-opto-mechanical devices are DPTs \cite{andrews14_bidir_effic_conver_between_microw_optic_light}.
To date, DPTs have demonstrated higher conversion efficiency \cite{higginbotham18_harnes_elect_optic_correl_effic_mechan_conver} while singly-parametric transducers have demonstrated larger bandwith operation \cite{hease20_bidir_elect_optic_wavel_conver}.
In this work we focus on DPTs whose action, when linearized about strong parametric optical and microwaves pumps, can be viewed as two-port Gaussian bosonic channels.
Furthermore, we only consider the on-resonance frequency mode in the steady-state and resolved-sideband limits, where the DPT channel implements either beamsplitter-type or two-mode squeezing-type interactions between the itinerant microwave and optical modes.

In order to find explicit thresholds on loss, noise, and cooperativity parameters under which DPTs can achieve quantum operation, we employ the Choi–Jamiołkowski isomorphism between states and channels \cite{holevo11_choi_jamiol_forms_quant_gauss_chann}, separability and positive partial transpose conditions for Gaussian states \cite{horodecki09_quant_entan, weedbrook12_gauss_quant_infor, giedke02_charac_gauss_operat_distil_gauss_states}, and entanglement-breaking conditions for one-mode Gaussian channels \cite{horodecki03_entan_break_chann, holevo08_entan_break_chann_infin_dimen}.
We start with a discussion of our DPT model and approximations in Section \ref{sec:deviceModel}.
Next, in Section \ref{sec:squeezing-two-mode}, we show that the two-mode squeezing-type interaction with vacuum inputs produces distillable entanglement for all DPT device parameters by relating it to a two-mode OE squeezed-lossy-state with effective squeezing and loss parameters.
Then, in Section \ref{sec:beamsplitter-two-mode}, for the beamsplitter-type two-mode channel, we find closed-form expressions for the DPT parameter thresholds which define regions where it is capable of producing OE distillable-entanglement, only capable of producing OE bound-entanglement, and never capable of entangling OE modes.
Finally, in Section \ref{sec:beamsplitter-one-mode}, we demonstrate that the entanglement-breaking thresholds for the simplest one-mode up-conversion and down-conversion channels are strictly worse than the two-mode thresholds.

\section{Doubly-Parametric Transducer Model \label{sec:deviceModel}}

\begin{figure*}
  \subfloat[\label{fig:couplings}\vspace{-15pt}]{\begin{tikzpicture}[
  on grid, node distance=6.5em, bend angle=25, scale=0.9,
  mode/.style={circle, draw=#1!50, fill=#1!20, thick, minimum size=4.4em},
  arrow/.style={->, shorten <=1pt, >={Stealth[round]}, thick, bend right},
  wave/.style={gray, thick, decorate, decoration={snake,amplitude=0.8, segment length=5.0}}
  ]
  \def\outenv{_{\text{out}}^{\text{e}}}
  \def\inenv{_{\text{in}}^{\text{e}}}
  \node[align=center,mode=red,               label={[text depth=0pt]above:optical}]   (a) {$\hat a$, $\hat a^\dagger$  \\[1mm] $\Delta_a$};
  \node[align=center,mode=green, right=of a, label={[text depth=0pt]above:mediator}]  (c) {$\hat c$, $\hat c^\dagger$ \\[1mm] $\omega_m$};
  \node[align=center,mode=blue,  right=of c, label={[text depth=0pt]above:microwave}] (b) {$\hat b$, $\hat b^\dagger$   \\[1mm] $\Delta_b$};
  \node at ($(a)!0.5!(c)$) {$G_a$};
  \node at ($(b)!0.5!(c)$) {$G_b$};
  \path (a.340) edge[arrow] (c.200) (c.160) edge[arrow] (a.20);
  \path (c.340) edge[arrow] (b.200) (b.160) edge[arrow] (c.20);
  \node[align=center,below=of a] (Ea) {$\hat a\outenv\quad\hat a\inenv$};
  \node[align=center,below=of b] (Eb) {$\hat b\outenv\quad\hat b\inenv$};
  \node[align=center,below=of c] (Ec) {$\hat c\outenv\quad\hat c\inenv$};
  \path (a) -- node {$\kappa_{a}^{\text{e}}$} (Ea);
  \path (b) -- node {$\kappa_{b}^{\text{e}}$} (Eb);
  \path (c) -- node {$\gamma_m$} (Ec);
  \node[below=0.6 of Ec] {environment};
  \draw[wave] ($(Ea.north)+(-2.2,0.3)$) -- ($(Eb.north)+(2.2,0.3)$);
  \path (a.250) edge[dashed,arrow,red  ] (Ea.125) (Ea.60) edge[dashed,arrow,red  ] (a.290);
  \path (b.250) edge[dashed,arrow,blue ] (Eb.115) (Eb.60) edge[dashed,arrow,blue ] (b.290);
  \path (c.250) edge[dashed,arrow,green] (Ec.125) (Ec.60) edge[dashed,arrow,green] (c.290);
  \node[align=center,left=1.8 of a ] (ain) {$\hat a_{\text{out}}^{\text{c}}$ \\[2mm] $\hat a_{\text{in}}^{\text{c}}$};
  \node[align=center,right=1.8 of b] (bin) {$\hat b_{\text{in}}^{\text{c}}$  \\[2mm] $\hat b_{\text{out}}^{\text{c}}$};
  \path (a) -- node {$\kappa_{a}^{\text{c}}$} (ain);
  \path (b) -- node {$\kappa_{b}^{\text{c}}$} (bin);
  \path (ain.320) edge[arrow,red ] (a.200)   (a.160)   edge[arrow,red ] (ain.40);
  \path (b.340)   edge[arrow,blue] (bin.220) (bin.140) edge[arrow,blue] (b.20);
\end{tikzpicture}}
  \subfloat[\label{fig:circuit}\vspace{-2pt}]{\begin{tikzpicture}
  [on grid, node distance=2em,
  R/.style={very thick, red},
  B/.style={very thick, blue},
  RD/.style={very thick, dashed, red},
  BD/.style={very thick, dashed, blue},
  G/.style={very thick, dashed, green},
  op/.style={draw, thick, black, fill=white},
  box/.style={draw, black, thick, dotted},
  pics/bs/.style args={nodes #1 #2 colors #3 #4 label #5 space #6 position #7}{code={
      \def\above{above}\def\argpos{#7}
      \coordinate (bsA0) at ($(#1)+(#6,0)$);
      \coordinate (bsB0) at ($(#2)+(#6,0)$);
      \coordinate (bsA1) at ($(bsB0.center)!1!\ifx\argpos\above \else - \fi 90:(bsA0.center)$);
      \coordinate (bsB1) at ($(bsA0.center)!1!\ifx\argpos\above - \else \fi 90:(bsB0.center)$) ;
      \draw[#3] (#1) -- (bsA0.center) -- (bsA1.center) -- ($(bsA1)+(#6,0)$);
      \draw[#4] (#2) -- (bsB0.center) -- (bsB1.center) -- ($(bsB1)+(#6,0)$);
      \coordinate (bsC0) at ($(bsA0)!0.5!(bsB0)$);
      \coordinate (bsC1) at ($(bsA1)!0.5!(bsB1)$);
      \draw[black, thick]  ($(bsC0)!0.25!(bsC1)$) -- ($(bsC0)!0.75!(bsC1)$)
      node[midway,label=\ifx\argpos\above below \else above \fi:#5] {};
      \coordinate (#1) at ($(bsB1)+(#6,0)$) {};
      \coordinate (#2) at ($(bsA1)+(#6,0)$) {};
    }},
  pics/op3/.style args={color #1 at #2 and #3 at #4 and #5 at #6 with space #7 label top #8 bottom #9}{code={
      \draw[#1] (#2) -- ++(#7,0) coordinate[midway] (op3a);
      \draw[#3] (#4) -- ++(#7,0);
      \draw[#5] (#6) -- ++(#7,0) coordinate[midway] (op3b);
      \node[fit={(op3a) (op3b)}] (T) {};
      \coordinate (#2) at ($(#2)+(#7,0)$) {};
      \coordinate (#4) at ($(#4)+(#7,0)$) {};
      \coordinate (#6) at ($(#6)+(#7,0)$) {};
      \node[below] (Ta) at (T.north) {#8};
      \node[above] (Tb) at (T.south) {#9};
      \node[op,inner xsep=0.4em,inner ysep=0.6em,fit={(op3a) (op3b) (Ta) (Tb)}] (T) {};
      \node[below] (Ta) at (T.north) {#8};
      \node[above] (Tb) at (T.south) {#9};
    }},
  ]
  \coordinate                                                (At);
  \coordinate[below=of At,label=left:$\ket{0}$]              (Ae);
  \coordinate[below=0.6 of Ae] (c);
  \coordinate[below=0.6 of c, label=left:$\ket{0}$]          (Be);
  \coordinate[below=of Be]                                   (Bt);
  \coordinate[label=left:$\ket{0}$] (Ae1) at ($(At)-(1.2,0)$);
  \coordinate[label=left:$\ket{0}$] (Be1) at ($(Bt)-(1.2,0)$);
  \coordinate[above left=1.0 of Ae1, xshift=-5, label=left:$\hat a_{\text{in}}$] (At1);  %
  \coordinate[below left=1.0 of Be1, xshift=-5, label=left:$\hat b_{\text{in}}$] (Bt1);  %
  \draw[R] (At1) -- (At1 -| Ae1) coordinate (At1);
  \draw[B] (Bt1) -- (Bt1 -| Be1) coordinate (Bt1);
  \node (c) at (c) {$\hat\rho(n_{th})$};
  \pic {bs={nodes At1 Ae1 colors R RD label $\delta_a$ space 0.2 position below}};
  \pic {bs={nodes Bt1 Be1 colors B BD label $\delta_b$ space 0.2 position above}};
  \draw[R] (Ae1) -- (At);
  \draw[B] (Be1) -- (Bt);
  \pic {bs={nodes At Ae colors R RD label $\tau_a$ space 0.15 position below}};
  \pic {bs={nodes Bt Be colors B BD label $\tau_b$ space 0.15 position above}};
  \draw[G] (c) -- (c -| At) coordinate (c);
  \pic {op3={color R at Ae and G at c and B at Be with space 1.0 label top $C_{a}$ bottom $C_{b}$}};
  \draw[RD] (At) -- (At -| Ae) coordinate (At);
  \draw[BD] (Bt) -- (Bt -| Be) coordinate (Bt);
  \pic {bs={nodes At Ae colors RD R label $\tau_a$ space 0.15 position below}};
  \pic {bs={nodes Bt Be colors BD B label $\tau_b$ space 0.15 position above}};
  \draw[G] (c) -- ++ (0.5,0)  coordinate (c);
  \draw[R] (At) --++(0.0, 0) coordinate (At);
  \draw[B] (Bt) --++(0.0, 0) coordinate (Bt);
  \node[box, fit={(-0.45,0.05) (3.3,-2.6)}] (box) {};
  \coordinate[above=of At, label=left:$\ket{0}$] (Ae1);
  \coordinate[below=of Bt, label=left:$\ket{0}$] (Be1);
  \pic {bs={nodes Ae1 At colors RD R label $\epsilon_a$ space 0.2 position below}};
  \pic {bs={nodes Be1 Bt colors BD B label $\epsilon_b$ space 0.2 position above}};
  \draw[R] (Ae1) -- ++(0.7, 0) node[black,anchor=west] {$\hat a_{\text{out}}$};
  \draw[B] (Be1) -- ++(0.7, 0) node[black,anchor=west] {$\hat b_{\text{out}}$};
  \node[draw, black,, fit={(-1.7,0.9) (4.2,-3.5)}] {};
\end{tikzpicture}}
  \caption{
    a) Diagram showing the couplings in the linearized operator equations of motion in a reference frame rotating at the pump frequencies.  Circles represent internal resonator modes.
    b) The equivalent circuit diagram derived from input-output relations of the central frequency mode in the resolved sideband limit.  In addition to the coupling losses arising from the equations of motion, this also includes the effect of transmission losses on the itinerant optical and microwave modes. Environmental modes are represented with dashed lines.
    Throughout this paper we use the convention of denoting optical, microwave, and mediating modes with red, blue, and green respectively.}
\end{figure*}
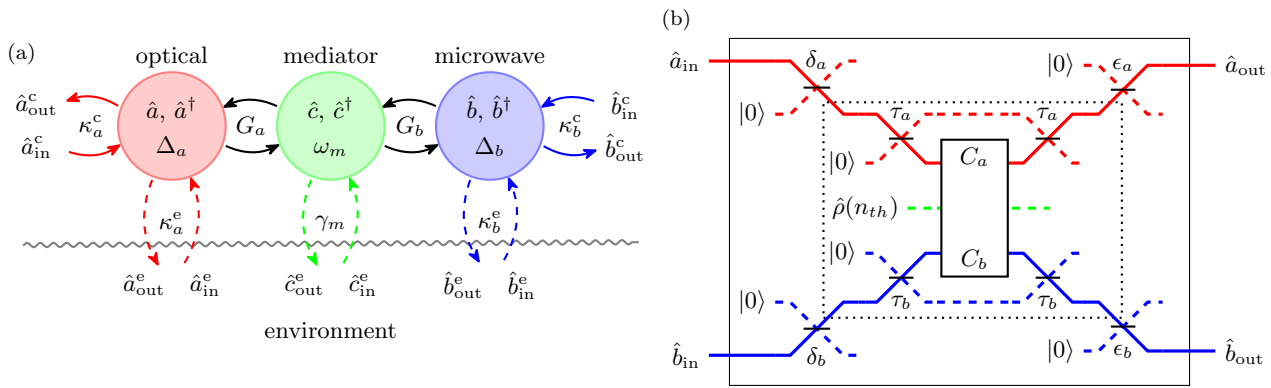

In general, the Hamiltonian describing a given physical DPT implementation will be nonlinear.
However, the relatively weak experimentally achievable bare coupling rates between the mediating mode and a single-photon can be enhanced with strong coherent optical and microwave pumps driving the doubly parametric interaction. In this regime, the operator equations of motion can be safely linearized about the strong pump fields \cite{aspelmeyer14_cavit_optom}.
Thus, as the starting point for our study, we take the following set of linear Heisenberg-Langevin equations of motion, which capture the essential couplings that most physical DPT implementations can be reduced to \cite{andrews14_bidir_effic_conver_between_microw_optic_light}
\begin{align}
  {\dot {\hat a}} & = \left(i\Delta_a - \frac{\kappa_a}{2}\right) {\hat a}
                                         + iG_a \left(\hat c + \hat c^\dagger\right)
                                         + \sqrt{\kappa_a^{\text{c}}} {\hat a_{\text{in}}^{\text{c}}}
                                         + \sqrt{\kappa_a^{\text{e}}} {\hat a_{\text{in}}^{\text{e}}}
  \nonumber \\
  {\dot {\hat b}} & = \left(i\Delta_b - \frac{\kappa_b}{2}\right) {\hat b}
                                         + iG_b \left(\hat c + \hat c^\dagger\right)
                                         + \sqrt{\kappa_b^{\text{c}}} {\hat b_{\text{in}}^{\text{c}}}
                                         + \sqrt{\kappa_b^{\text{e}}} {\hat b_{\text{in}}^{\text{e}}}
  \nonumber \\
  {\dot {\hat c}} & = -\left(i\omega_m + \frac{\gamma_m}{2}\right) {\hat c}
                                         + iG_a \left(\hat a + \hat a^\dagger\right)
  \nonumber \\
                         &\hspace{8.6em} + iG_b \left(\hat b + \hat b^\dagger\right)
                                         + \sqrt{\gamma_m} {\hat c_{in}} \label{eqn:eqms}.
\end{align}

Fig. \ref{fig:couplings} illustrates the couplings in these equations of motion, where the annihilation operators $\hat a$ and $\hat b$ refer to the optical and microwave cavity modes, respectively, while $\hat c$ refers to the mediating mode which couples to $\hat a$ and $\hat b$ at parametrically enhanced coupling rates of $G_a$ and $G_b$, respectively.
In the linearized regime, it is convenient to work in a frame rotating at the pump frequencies, thus $\Delta_a$ and $\Delta_b$ describe the frequencies of $\hat a$ and $\hat b$ in terms of the detuning from their respective pump frequencies.

As these devices are not isolated systems, we use input-output theory in the above expressions to relate the internal operators to propagating and environmental operators \cite{clerk10_introd_to_quant_noise_measur_amplif}.
We assume the optical, microwave, and mediating modes couple to experimentally inaccessible environmental bath modes $\hat a_{\{\text{in,out}\}}^{\text{e}}$, $\hat b_{\{\text{in,out}\}}^{\text{e}}$, and $\hat c_{\{\text{in,out}\}}^{\text{e}}$ at rates $\kappa_a^{\text{e}}$, $\kappa_b^{\text{e}}$, and $\gamma_m$, respectively, while the optical and microwave cavities couple to itinerant input and output modes $\hat a_{\{\text{in,out}\}}^{\text{c}}$ and $\hat b_{\{\text{in,out}\}}^{\text{c}}$ at rates $\kappa_a^{\text{c}}$ and $\kappa_b^{\text{c}}$, respectively.
Thus the optical and microwave cavities have linewidths of $\kappa_a = \kappa_a^{\text{c}}+\kappa_a^{\text{e}}$ and $\kappa_b = \kappa_b^{\text{c}}+\kappa_b^{\text{e}}$, respectively.
We assume the device is cooled to a temperature where the thermal occupancy of the environmental baths are negligible at the optical and microwave frequencies, however the environment of the typically lower-frequency mediating mode is taken to have a thermal occupancy of $n_{\text{th}}$ bosons.
After accounting for all these couplings, we solve for the steady-state in the frequency domain which results in a transfer function $\boldsymbol{\Xi}(\omega)$ describing a unitary transformation on all itinerant frequency modes (input/output modes plus environmental modes).
As expected from the assumed linear dynamics, $\boldsymbol{\Xi}(\omega)$ represents a symplectic transform with terms describing beamsplitter-type and squeezing-type interactions between the optical and microwave input/output modes.

The full model contains an unwieldy number of parameters, but with several approximations the transducer can be characterized by just five parameters.
First, we only consider the itinerant optical and microwave sideband frequency modes that are on resonance with the mediating mode ($\boldsymbol{\Xi}(\omega=\pm\omega_m)$ in a frame rotating about the strong pumps).
The interaction of these sidebands with the mediating mode is maximized by setting the magnitude of the relative pump detunings $\Delta_{\{a,b\}}$ to be equal to the mediating resonator's frequency, after accounting for frequency shifts due to coupling.
We will refer to setting a pump below its cavity resonance frequency ($\Delta_{\{a,b\}} = - \omega_m$, maximizing anti-Stokes scattering) as \emph{red} detuned, while setting it above ($\Delta_{\{a,b\}} = + \omega_m$, maximizing Stokes scattering) is \emph{blue} detuned.
Next, we assume the mediating resonator is high-Q and we make resolved sideband approximations so that $4\omega_m \gg \kappa_a, \kappa_b, \gamma_m$.
This allows us to neglect the Stokes sideband when red detuned and neglect the anti-Stokes sideband when blue detuned.
Thus, after these approximations, we only consider itinerant signal modes that are on resonance with their respective resonators.
We then define two cooperativities as $C_i = \frac{4 G_i^2}{\kappa_i \gamma_m}$ where $i\in\{a,b\}$ which capture the rate at which information may be exchanged between the optical or microwave mode and the mediating mode relative to their cavity decay rates.
We also define $\tau_i = \frac{\kappa^c_i}{\kappa_i}$ where $i\in\{a,b\}$ as the cavity coupling transmissivities.
Thus, we now have the five dimensionless parameters, $C_{\{a,b\}}$, $\tau_{\{a,b\}}$, and $n_{\text{th}}$, which characterize a DPT.

Under these approximations, the coupling transmissivity $\tau_i$, which describes how over or under coupled the itinerant field is to the cavity field, can be modeled by an environmental mode initialized in vacuum mixing with the input and then output modes via beamsplitters, each with the same transmissivity $\tau_i$, as shown in Fig. \ref{fig:circuit}.
We also include parameters describing losses on the input and output modes that would be ``external'' to the cavities.
Such losses might include transmission losses, imperfect mode matching between cavity and detector modes, or any other losses that can be described by an effective beamsplitter with an environmental vacuum mode as shown in Fig. \ref{fig:circuit} where $\delta_i$ and $\epsilon_i$ refer to the beamsplitter transmissivities before and after the DPT respectively, while the subscript $i\in\{a,b\}$ refers to whether it is on the optical or microwave mode respectively.

In order to explicitly trace-out the inaccessible environmental modes, we reduce the unitary $\boldsymbol{\Xi}(\omega=\pm\omega_m)$ to a two-mode Gaussian channel acting on OE modes. This channel can be explicitly described by two $4\times 4$ matrices, a unitary-like $\boldsymbol{T}$ and an added noise-like $\boldsymbol{N}$, that describe the action of the channel on an arbitrary covariance matrix as $\boldsymbol{V}\to \boldsymbol{T}\boldsymbol{V}\boldsymbol{T}^\top + \boldsymbol{N}$ \cite{weedbrook12_gauss_quant_infor,eisert05_gauss_quant_chann} (see Appendix \ref{sec:in-out} for the explicit forms). Note that we have chosen the convention of vacuum variance corresponding to $1/2$ quanta for the single frequency modes considered under our approximations.

\section{Squeezing-type Interaction}

\subsection{Two-Mode\label{sec:squeezing-two-mode}}

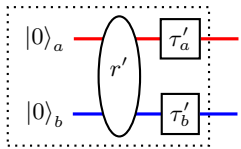
\begin{figure}
  \begin{tikzpicture}
  [on grid,
  R/.style={very thick, red},
  B/.style={very thick, blue},
  G/.style={very thick, green},
  op/.style={draw, thick, black, fill=white},
  box/.style={draw, black, thick, dotted},
  ]
  \node (a)          {$\ket{0}_{a}$};
  \node[below=of a] (b) {$\ket{0}_{b}$};
  \draw[R] (a) -- ++(1.8,0) node[op] (af) {$\tau'_{a}$};
  \draw[B] (b) -- ++(1.8,0) node[op] (bf) {$\tau'_{b}$};
  \node[op,inner xsep=0,inner ysep=0.2em,ellipse,
  fit={($(a)!0.6!(af)$) ($(b)!0.5!(bf)$)}, text width=1.2em] (sqt) {$r'$};
  \node[box, fit={(a) (b) (sqt) (af) (bf)}] (box) {};
  \draw[R] (af.east) -- ++(0.5,0) coordinate (a);
  \draw[B] (bf.east) -- ++(0.5,0) coordinate (b);
\end{tikzpicture}
  \caption{\label{fig:sls} Effective two-mode squeezed lossy state describing the state generated by a DPT under a squeezing-type interaction with vacuum inputs. The variable $r'$ is the effective squeezing parameter while $\tau'_a$ and $\tau'_b$ are the effective single mode transmissivities.}
\end{figure}

Doubly-parametric transducers affect a two-mode squeezing-type interaction between OE modes when red detuning the microwave pump while blue detuning the optical pump, or vice versa.
When operated in this way, DPTs are always capable of producing distillable entanglement between OE modes.
While many different OE input states can result in distillable entanglement under this interaction, it is sufficient to consider the simplest case of vacuum inputs in order to demonstrate distillable entanglement for any possible DPT parameter values.
When both OE inputs are initialized with vacuum, the covariance matrix of the resulting output state $\hat\psi$ is found by computing $\boldsymbol{V} = \boldsymbol{T} (\boldsymbol{I}_4/2) \boldsymbol{T}^\top + \boldsymbol{N}$ where $\boldsymbol{I}_4$ is the $4\times 4$ identity matrix.
A break down in our linearization about a strong pump approximation occurs when $C_i - C_j = 1$ due to terms in $\boldsymbol{V}$ becoming infinite ($i,j = a,b$ when the optical pump is blue detuned and the microwave pump is red detuned, while $i,j = b,a$ in the opposite case).
Therefore we exclude the region near the $C_i - C_j = 1$ pole from the following discussion of DPTs operating as squeezers.
In this case of $1\times 1$ bipartite Gaussian states, the positive partial transpose (PPT) criteria is both a necessary and sufficient condition for distillable entanglement \cite{weedbrook12_gauss_quant_infor, horodecki09_quant_entan, giedke01_entan_criter_all_bipar_gauss_states}.
We find that for all parameters, $\hat\psi$ is entangled, and that the amount of entanglement quantified by the logarithmic negativity in terms of DPT parameters is given by
\begin{align}
     E_{N} &= -\ln \left(1 + 4\frac{X^{(1)} - \sqrt{X^{(0)}X^{(2)}}}
             {\left(1 - C_i + C_j\right)^2}\right)
\end{align}
where for compactness we introduce the quantity
\begin{align*}
   X^{(k)} &= C_i\epsilon_i\tau_i \left(C_j + n_{\text{th}} + 1\right)^k
                 + C_j\epsilon_j\tau_j  \left(C_i + n_{\text{th}}\right)^k .
\end{align*}

We can gain additional insight into the nature of the OE entanglement by relating $\hat\psi$ to an effective two-mode squeezed lossy state (TMSLS) as shown in Fig. \ref{fig:sls}.
A TMSLS has a specific form of the covariance matrix, which $\boldsymbol{V}$ always satisfies.
Therefore we solve for the TMSLS parameters of effective squeezing $r'$ and effective OE transmissivities $\tau'_{\{a,b\}}$ respectively in terms of the DPT parameters, finding
\begin{align}
  \cosh(r') &= \frac{8
                \left(C_i + n_{\text{th}}\right)
                \left(C_j + n_{\text{th}} + 1\right)}{
                \left(C_i - C_j - 1\right)^{2}} + 1 , \\
  \tau'_i &= \frac{C_i \tau_i\epsilon_i}{C_i + n_{\text{th}}} , \\ \text{and}\quad
  \tau'_j &= \frac{C_j \tau_j\epsilon_j}{C_j + n_{\text{th}} + 1} .
\end{align}

For all nonzero $C_{\{a,b\}}$, $\tau_{\{a,b\}}$, $\epsilon_{\{a,b\}}$, and $n_{\text{th}}$ we see that $r' > 0$ and $0 < \tau'_a, \tau'_b < 1$ which illustrates that this state is always entangled and mixed \cite{hoelscher-obermaier11_optim_gauss_entan_swapp}, and reaffirms the robustness of any TMSLS to single mode losses \cite{kotler20_tomog_entan_macros_mechan_objec}.
Therefore, in principle entanglement should be present even with low cooperativities and a hot mediating environment.
However, in transforming from device parameters to effective TMSLS parameters, we see that for large $n_{\text{th}}$, $\tau'_{\{a,b\}}$ scale with $n_{\text{th}}^{-1}$, while conversely, $r'$ scales with $2\ln\left(n_{\text{th}}\right)$.
So a device with $C_{\{a,b\}} \ll n_{\text{th}}$ would produce highly mixed entanglement from large squeezing followed by large losses.
Such a state would be very sensitive to any subsequent added noise, making it poorly suited for use within a larger network\cite{holevo08_entan_break_chann_infin_dimen}.

While this mode of operation is promising for demonstrating OE entanglement, extracting more useful, purer entangled states from this mixed OE state will require applying distillation or purification protocols which may prove difficult for this state.
As it is impossible to distill Gaussian states using only Gaussian operations, non-Gaussian measurements will be needed \cite{eisert02_distil_gauss_states_with_gauss, giedke02_charac_gauss_operat_distil_gauss_states, duan00_entan_purif_gauss_contin_variab_quant_states}.
However, this state is not easily compatible with single photon detection due to the bright pumps, which usually have small fractional frequency separation from the entangled modes due to the mediating mode's resonance frequency being low relative to the optical and microwave frequencies.
This poses the difficult experimental challenge of creating filters capable of rejecting the bright pumps.

\subsection{Single-Mode}
\begin{figure}
  \subfloat[\label{fig:downconvert}\vspace{-12pt}]{\hspace{12pt}\begin{tikzpicture}
  [on grid,
  R/.style={very thick, red},
  B/.style={very thick, blue},
  G/.style={very thick, green},
  BL/.style={very thick, black},
  op/.style={draw, thick, black, fill=white},
  box/.style={draw, black, thick, dotted},
  pics/one mode/.style args={start #1 space #2 and #3 label #4 colors #5 and #6}{code={
      \draw[#5] (#1) -- ++(#2,0) coordinate (#1);
      \node[op,right=0 of #1] (a0) {$\tau'_{\text{#4}}$};
      \draw[BL] (a0.east) -- ++(0.2,0) coordinate (#1);
      \node[op,right=0 of #1] (af) {$n'_{\text{#4}}$};
      \draw[#6] (af.east) -- ++(#3,0) coordinate (#1);
      \node[draw, black, thick, dotted, fit={(a0) (af)}] (box) {};
    }},
  ]
  \coordinate (a) at (0,0);
  \pic {one mode={start a space 1.0 and 1.0 label d colors R and B}};
  \node at (0.0,0.2) {optical};
  \node at (3.5,0.2) {microwave};
  \node at ($(box.north)+(0,0.2)$) {downconvert};
\end{tikzpicture}}

  \subfloat[\label{fig:upconvert}\vspace{-12pt}]{\begin{tikzpicture}
  [on grid,
  R/.style={very thick, red},
  B/.style={very thick, blue},
  G/.style={very thick, green},
  BL/.style={very thick, black},
  op/.style={draw, thick, black, fill=white},
  box/.style={draw, black, thick, dotted},
  pics/one mode/.style args={start #1 space #2 and #3 label #4 colors #5 and #6}{code={
      \draw[#5] (#1) -- ++(#2,0) coordinate (#1);
      \node[op,right=0 of #1] (a0) {$\tau'_{\text{#4}}$};
      \draw[BL] (a0.east) -- ++(0.2,0) coordinate (#1);
      \node[op,right=0 of #1] (af) {$n'_{\text{#4}}$};
      \draw[#6] (af.east) -- ++(#3,0) coordinate (#1);
      \node[draw, black, thick, dotted, fit={(a0) (af)}] (box) {};
    }},
  ]
  \coordinate (a) at (0.2,0);
  \pic {one mode={start a space 1.0 and 1.0 label u colors B and R}};
  \node at (0.0,0.2) {microwave};
  \node at (3.5,0.2) {optical};
  \node at ($(box.north)+(0,0.2)$) {upconvert};
\end{tikzpicture}\hspace{12pt}}
  
  \caption{\label{fig:single-mode}
    a) Single-mode downconversion channel obtained by initializing the microwave port with vacuum and tracing out the optical output in Fig. \ref{fig:circuit}.
    b) Single-mode upconversion channel obtained by initializing the optical port with vacuum and tracing out the microwave output in Fig. \ref{fig:circuit}.
    These single-mode channels can be completely characterized by effective transmissivity $\tau'_{\{\text{u},\text{d}\}}$ and added noise $n'_{\{\text{u},\text{d}\}}$ parameters.
  }
\end{figure}
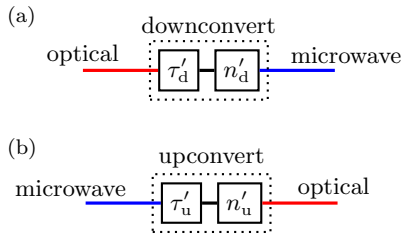
With the application of exchanging quantum information between optical and microwave frequencies in mind, an intuitive way to use a transducer is as a converter, which we define to be a one-mode channel where the input and output are at different frequencies.
There are two converter classifications distinguished by the direction of information flow: 1) optical-to-microwave downconversion (Fig. \ref{fig:downconvert}) and 2) microwave-to-optical upconversion (Fig. \ref{fig:upconvert}).
There are several ways to reduce the two-mode transducer considered thus far down to one input and one output \cite{zhang18_quant_trans_with_adapt_contr, zeuthen20_figur_merit_quant_trans, barzanjeh12_rever_optic_to_microw_quant_inter}.
Here we examine the simplest case: initialize one mode in vacuum and trace the other mode at the output.
When implementing the two-mode squeezing-type interaction, the converters become phase-insensitive phase-conjugating amplifiers (or attenuators if the loss exceeds the gain) \cite{jiang20_effic_bidir_piezo_optom_trans}.
In this configuration we find that the squeezing-type converters are always entanglement breaking \cite{holevo08_entan_break_chann_infin_dimen}, equivalent to a measure-and-prepare channel \cite{horodecki03_entan_break_chann}, have zero quantum capacity \cite{lercher13_stand_super_activ_gauss_chann_requir_squeez, horodecki09_quant_entan}, and the p-function of the output state is always positive \cite{ivan13_noncl_break_is_same_as}.
However, the squeezing-type converters can function as excellent amplifiers and, for example, may be useful for amplifying classical information or reading out a microwave qubit optically.

\section{Beamsplitter-type Interaction}
\subsection{Two-Mode\label{sec:beamsplitter-two-mode}}
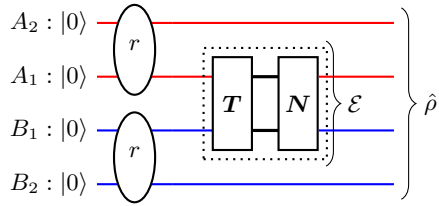
\begin{figure}
  \begin{tikzpicture}
  [on grid, node distance=2.2em,
  R/.style={thick, red},
  B/.style={thick, blue},
  G/.style={thick, green},
  BL/.style={very thick, black},
  op/.style={draw,thick,fill=white},
  pics/sq/.style args={color #1 between #2 and #3 with label #4 and space #5}{code={
      \draw[#1] (#2) -- ++(#5,0) coordinate[midway] (sqa);
      \draw[#1] (#3) -- ++(#5,0) coordinate[midway] (sqb);
      \node (sqt) at ($(sqa)!0.5!(sqb)$) {#4};
      \node[op,inner xsep=0,inner ysep=0.2em,ellipse,fit={(sqa) (sqb) (sqt)}] {#4};
      \coordinate (#2) at ($(#2)+(#5,0)$);
      \coordinate (#3) at ($(#3)+(#5,0)$);
    }},
  ]
  \coordinate[            label=left:$A_2:\ket{0}$] (A2);
  \coordinate[below=of A2,label=left:$A_1:\ket{0}$] (A1);
  \coordinate[below=of A1,label=left:$B_1:\ket{0}$] (B1);
  \coordinate[below=of B1,label=left:$B_2:\ket{0}$] (B2);
  \pic {sq={color R between A2 and A1 with label $r$ and space 1}};
  \pic {sq={color B between B1 and B2 with label $r$ and space 1}};
  \draw[R] (A1) --++(0.8,0) coordinate (A1);
  \draw[B] (B1) --++(0.8,0) coordinate (B1);
  \node[op, fit={(A1) (B1)}, inner sep=0.8em] (T) {};
  \node at ($(A1)!0.5!(B1)$) {$\boldsymbol{T}$};
  \draw[BL] (A1 -| T.east) -- ++(0.6,0) coordinate (A1);
  \draw[BL] (B1 -| T.east) -- ++(0.6,0) coordinate (B1);
  \node[op, fit={(A1) (B1)}, inner sep=0.8em] (N) {};
  \node at ($(A1)!0.5!(B1)$) {$\boldsymbol{N}$};
  \draw[R] (A1 -| N.east) -- ++(1.0,0) coordinate (A1);
  \draw[B] (B1 -| N.east) -- ++(1.0,0) coordinate (B1);
  \node[draw, black, thick, dotted, fit={(T) (N)}] {};
  \draw[R] (A2) -- (A2 -| A1) coordinate (A2);
  \draw[B] (B2) -- (B2 -| B1) coordinate (B2);
  \draw[decorate,decoration={brace,amplitude=0.6em,raise=0.3em}] ($(A2)+(0,0.2)$)
  -- node[xshift=1.5em,midway] {$\hat\rho$}           ($(B2)-(0,0.2)$);
  \draw[decorate,decoration={brace,amplitude=0.6em,raise=0.3em}] ($(N.north east)+(0,0.2)$)
  -- node[xshift=1.5em,midway] {$\mathcal{E}$}           ($(N.south east)-(0,0.2)$);
\end{tikzpicture}
  \caption{\label{fig:two-mode}
    Given the channel $\mathcal{E}$, the Choi–Jamiołkowski isomorphism gives the dual state $\hat\rho$, which is found by $\mathcal{E}$ acting on inputs which are each maximally entangled with an ancilla.
    For continuous variable systems, the maximally entangled states are infinitely squeezed two-mode squeezed vacuum states ($r\to\infty$).
    The dashed box represents the two-mode channel between optical and microwave modes that the transducer implements.
  }
\end{figure}

\begin{figure*}
    \input{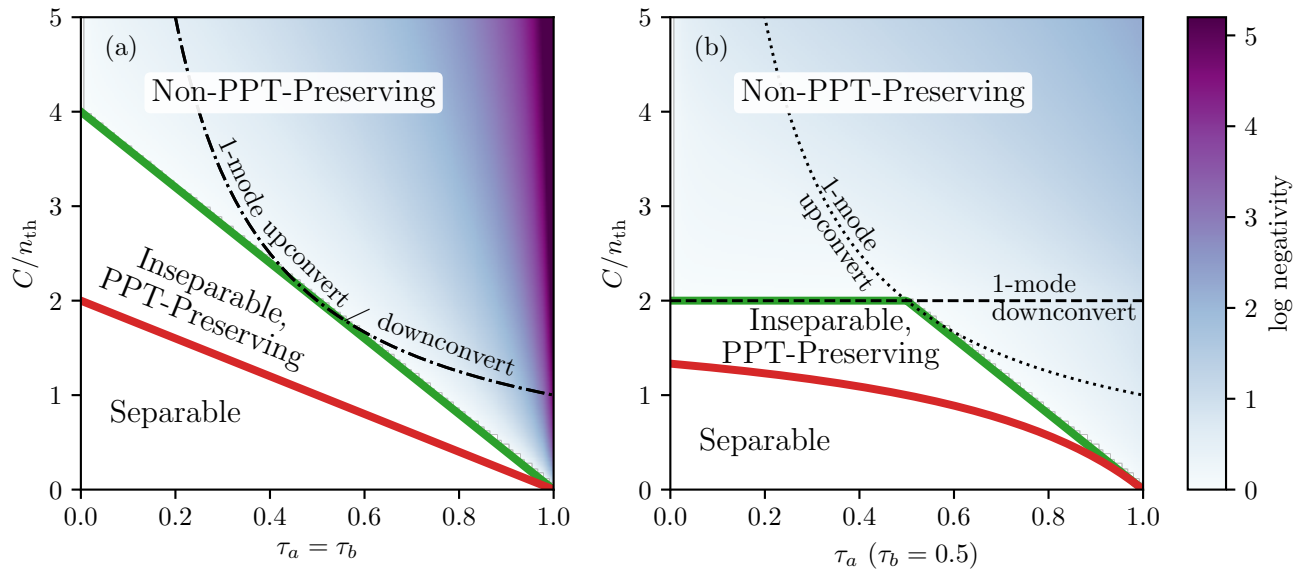}
  \caption{\label{fig:thresholdPlots}
    Plots of entanglement thresholds for the beamsplitter-type DPT interaction with equal cooperativites ($C = C_a = C_b$) and no transmission losses ($\delta_a = \delta_b = \epsilon_a = \epsilon_b = 1$), in which case the thresholds can be characterized simply by the quantum cooperativity $C/n_{\text{th}}$ which is related to the average thermal occupation of the mediating mode due to radiation pressure cooling \cite{aspelmeyer14_cavit_optom}.
    a) Thresholds as a function of equal optical and microwave coupling transmissivities.
    b) Thresholds as a function of only optical transmissivity with microwave transmissivity fixed at 0.5 to illustrate the different functional behavior between the thresholds.
    For both plots, below the red curve the two-mode transducer channel is separable (Eq.~\eqref{eqn:separability}), while between the red and green curves the channel is capable of producing bound-entanglement, and above the green curve the channel is capable of producing distillable entanglement (Eq.~\eqref{eqn:ppt}). The entanglement breaking thresholds for the one-mode upconversion (Eq.~\eqref{eqn:upconvert}) and downconversion (Eq.~\eqref{eqn:downconvert}) channels are indicated with the dashed curves and are never better than the two-mode transducer channel thresholds.
  }
\end{figure*}

Doubly-parametric transducers affect a two-mode beamsplitter-type interaction between OE modes when both pumps are red detuned.
We want to know the most general conditions under which this two-mode channel can establish OE entanglement, allowing arbitrary (potentially non-Gaussian) input states with ancilla modes, operations, and measurements, all of which are separable across the optical-microwave partition.
Since this interaction is not capable of intrinsically generating entanglement, it cannot entangle its OE outputs by simply initializing the inputs with vacuum as with the two-mode squeezing-type interaction.
Therefore we must use a more general method for understanding the conditions under which a quantum channel is either separable or capable of entangling the modes it acts on.

In order to find such conditions, we use the Choi–Jamiołkowski (CJ) isomorphism to find the four-mode Choi state $\hat{\rho}$ which is dual to the two-mode transducer channel $\mathcal{E}$ \cite{holevo11_choi_jamiol_forms_quant_gauss_chann}.
To calculate $\hat\rho$, we will first define $A_1$ and $B_1$ as the respective optical and microwave modes that $\mathcal{E}$ acts on, while $A_2$ and $B_2$ are single-mode optical and microwave CV ancilla modes respectively.
Both the $A_1$-$A_2$ and $B_1$-$B_2$ systems are then initialized in infinitely squeezed two-mode squeezed vacuum states, and after the channel $\mathcal{E}$ has acted on the $A_1$ and $B_1$ modes we obtain $\hat\rho$, as illustrated in Fig. \ref{fig:two-mode} (see Appendix \ref{sec:ppt-threshold} for their explicit expressions). Since $\mathcal{E}$ is Gaussian, $\hat\rho$ is completely characterized by a covariance matrix, which makes the following calculations tractable.

The CJ isomorphism tells us that $\mathcal{E}$ is separable if and only if $\hat{\rho}$ is separable with respect to the bipartition defined by the $A$-$B$ systems.
Conversely, $\mathcal{E}$ can facilitate entanglement creation across the $A$-$B$ partition if and only if $\hat{\rho}$ is inseparable with respect to the $A$-$B$ partition \cite{giedke02_charac_gauss_operat_distil_gauss_states, cirac01_entan_operat_their_implem_using}.
So we use the covariance matrix $\boldsymbol{V}$ of the Choi state to find when $\mathcal{E}$ is separable.  By implementing the iterative method for finding general Gaussian bipartite separability for an arbitrary numbers of modes \cite{giedke01_entan_criter_all_bipar_gauss_states}, we numerically fit the following threshold, finding $\mathcal{E}$ is separable with respect to the $A$-$B$ partition if and only if
\begin{align}
  n_{\text{th}} \geq \nu_a + \nu_b
  \label{eqn:separability}
\end{align}
where
\begin{align}
  \nu_i = \frac{\delta_i \tau_i C_i}{1 - \delta_i\epsilon_i(1 - 2\tau_i)^2}
\end{align}

The CJ isomoprhism also relates the positivity after partial transposition (PPT) criteria between a channel and its dual state.
By definition, the Choi state $\hat{\rho}$ is PPT with respect to the $A$-$B$ partition if and only if $\hat{\rho}^{\top_A}\ge 0$.
Similarly we call the channel $\mathcal{E}$ PPT-preserving with respect to the $A$-$B$ partition when for all states $\hat\rho_{\text{in}}^{\top A}\ge 0$, we have $\mathcal{E}(\hat\rho_{\text{in}})^{\top_A}\ge 0$.
Then by the CJ isomoprhism we know that $\mathcal{E}$ is PPT-preserving if and only if $\hat{\rho}$ is PPT, both with respect to the $A$-$B$ partition (see Appendix \ref{sec:ppt-proof}) \cite{giedke02_charac_gauss_operat_distil_gauss_states, cirac01_entan_operat_their_implem_using, khatri20_princ_quant_commun_theor}.
So by checking the PPT criteria on $\boldsymbol{V}$ (see Appendix \ref{sec:ppt-threshold}) we find that $\mathcal{E}$ is PPT-preserving with respect to the $A$-$B$ partition if and only if
\begin{align}
  n_{\text{th}} \geq \max \left\{ \nu_a, \nu_b \right\}
  \label{eqn:ppt}
\end{align}
Plots of these thresholds for specific parameter regions are shown in Fig. \ref{fig:thresholdPlots}.

As all separable states are PPT and thus all separable channels are PPT-preserving, the above thresholds given in Eqs.~\eqref{eqn:separability} and \eqref{eqn:ppt} define three distinct regions for the two-mode beamsplitter-type DPT channel:
\renewcommand{\labelenumi}{(\roman{enumi})}%
\begin{enumerate}
\item \textbf{Separable}:
  $\mathcal{E}$ is separable and thus cannot be used to entangle OE modes if and only if Eq.~\eqref{eqn:separability} is true.
  Since in this region the state of the optical and microwave modes output from $\mathcal{E}$ will never be entangled, they can always be prepared using local operations on the OE modes and classical communication between them.
\item \textbf{Inseparable, PPT-Preserving}:
  $\mathcal{E}$ is Inseparable and PPT-Preserving if and only if Eq.~\eqref{eqn:ppt} is true and Eq.~\eqref{eqn:separability} is false.
  In this region $\mathcal{E}$ is capable of outputting bound entangled OE states (it is, of course, still possible for separable states to be output).
  In order to create bound entangled Gaussian states, the input to $\mathcal{E}$ must be locally entangled with ancillas on both the microwave and optical sides of the channel because $1\times N$ bipartite Gaussian systems cannot be bound entangled \cite{werner01_bound_entan_gauss_states}.
  However it may be possible to create bound entangled non-Gaussian states without the use of ancilla modes.
  Additionally, in this region the beamsplitter DPT is not capable of generating entanglement which can be transferred to $1\times 1$ OE qubit systems (since $1\times 1$ qubit entanglement is always non-PPT \cite{horodecki97_insep_two_spin_matric_can} and LOCC is PPT preserving \cite{horodecki09_quant_entan}.
\item \textbf{Non-PPT-Preserving}:
    $\mathcal{E}$ is Non-PPT-Preserving if and only if Eq.~\eqref{eqn:ppt} is false.
    In this region $\mathcal{E}$ is capable of outputting distillably entangled OE modes (it is, of course, still possible for separable and bound entangled states to be output).
\end{enumerate}

Devices capable of achieving high cooperativity and low loss on either the optical or microwave side, but not both, are still capable of crossing into quantum operation despite the poorer performing side. This is a direct consequence of the Non-PPT-preserving threshold given by Eq.~\eqref{eqn:ppt} being a maximum of two expressions that each involve parameters strictly on either the optical or microwave side.
However, designing experiments capable of utilizing both of the two ports of the beamsplitter-type channel will be more challenging than simply ignoring the unused input and output ports when operating as single-mode up or down conversion channels.

\subsection{Single-Mode\label{sec:beamsplitter-one-mode}}

We again examine the performance of a transducer when used as a single-mode converter, but now for the case when both pumps are red detuned.
As before, one input is eliminated by initializing it in vacuum and one output is eliminated by tracing the output.
Now the upconverter and downconverter are attenuation channels (Fig. \ref{fig:single-mode}), which are completely characterized by an effective power loss $\tau_{\text{u/d}}'$ and added noise photons at the output $n_{\text{u/d}}'$ given by
\begin{align}
    \tau'_{\{u,d\}} &= \frac{4\delta_{\{b,a\}}\epsilon_{\{a,b\}}\tau_a\tau_b C_a C_b}{(1+C_a+C_b)^2} \label{eqn:eff-loss}
    \\
    n'_{\{u,d\}} &= \frac{1}{2} + \frac{2\epsilon_{\{a,b\}} \tau_{\{a,b\}} C_{\{a,b\}}(2n_{\text{th}} - \delta_{\{a,b\}} \tau_{\{a,b\}} C_{\{a,b\}}) }{(1+C_a+C_b)^2} \label{eqn:eff-noise}
\end{align}
As expected, the upconversion and downconverersion channels both become the identity channel in the limit $n_{\text{th}}\to 0$, $C_a=C_b \to \infty$, and $\tau_{\{a,b\}},\delta_{\{a,b\}},\epsilon_{\{a,b\}} = 1$, and are not entanglement breaking if and only if
\begin{align}
    \label{eqn:upconvert}
    \text{Upconversion:}   \quad & n_{\text{th}} < \tau_b \delta_b C_b \\
    \label{eqn:downconvert}
    \text{Downconversion:} \quad & n_{\text{th}} < \tau_a \delta_a C_a
\end{align}
These thresholds follow directly from the effective loss and noise parameters (Eqs.~\eqref{eqn:eff-loss} and \eqref{eqn:eff-noise}) along with the entanglement breaking criteria for one-mode channels \cite{holevo08_entan_break_chann_infin_dimen}.
If these inequalities are violated, the converters are equivalent to a measure-and-prepare channel \cite{horodecki03_entan_break_chann}.
Therefore, Eqs.~\eqref{eqn:upconvert} and \eqref{eqn:downconvert} are also necessary conditions for non-zero quantum capacity \cite{lercher13_stand_super_activ_gauss_chann_requir_squeez}.
These thresholds are more restrictive than those in the previous section (Eqs.~\eqref{eqn:ppt} and \eqref{eqn:separability}) which can be seen in Fig. \ref{fig:thresholdPlots}.
This is partially due to the fact that tracing over the unused output discards useful information \cite{higginbotham18_harnes_elect_optic_correl_effic_mechan_conver}.
Additionally, it is interesting to note that loss incurred after added noise due to transduction does not affect entanglement breaking thresholds ($\epsilon_{\{a,b\}}$ does not appear in Eqs.~\eqref{eqn:upconvert} and \eqref{eqn:downconvert}), illustrating the non-commutativity of these two sources of decoherence.
Thus, it is possible to operate a transducer in a parameter regime where upconversion is entanglement breaking while downconversion is not, or vice versa (see Fig. \ref{fig:thresholdPlots}b).
Finally, higher cooperativities are better for moving into the quantum regime, but once the converters are capable of quantum operation it may be necessary to strike a balance between large and equal cooperativities so that $\tau_{\{u,d\}}$ is not too small.

\section{Conclusion}

We have found explicit thresholds on DPT parameters that must be satisfied in order for the two-mode beamsplitter-type channel to be capable of distillably entangling OE modes.
In contrast, we have shown that the two-mode squeezing-type interaction with vacuum inputs always creates distillable OE entanglement.
Thus, we conclude that for low cooperativity and ``hot'' DPTs, the squeezing-type interaction is the only viable way for such devices to achieve quantum operation.
However, technical challenges limit the ability to take advantage of the highly mixed entangled states produced by such devices.
As DPT device performance improves to below the beamsplitter-type distillable entanglement threshold, it may be that using devices in such a configuration offers benefits over the squeezing-type interaction.
For example, using non-Gaussian inputs to the beamsplitter-type interaction may allow the output OE entanglement to be distilled using only Gaussian operations.

Current quantum information devices are based on a diverse set of physical platforms.
To continue to build ever more complicated and powerful systems, we need to leverage the strengths of disparate systems.
Quantum transducers capable of coherently linking disparate frequency modes will be essential components of future heterogeneous quantum networks.
Yet, any transducer will introduce loss and noise into the system, which degrades their utility.
Entanglement distillation, purification, concentration, or error correction protocols can help overcome the transduction imperfections;
however, the entanglement thresholds found in this work must be overcome before a transducer is ready to support such protocols or be integrated into a quantum network.

\section*{Acknowledgements}
The authors would like to thank Emanuel Knill, Scott Glancy, and Sristy Agrawal for helpful suggestions and discussions.
At the time this work was performed, C. Rau, A. Kyle, A. Kwiatkowski, and E. Shojaee were supported as Associates in the Professional Research Experience Program (PREP) operated jointly by NIST and the University of Colorado Boulder under Award no. 70NANB18H006 from the U.S. Department of Commerce.
This is a contribution of the National Institute of Standards and Technology, not subject to U.S. copyright.

\appendix
\section{From Input-Output Relations to Two-Mode Gaussian Channels  \label{sec:in-out}}

We start by writing the equations of motion as given in Eqs.~\eqref{eqn:eqms} as a state space model
\begin{align}
\dot{\boldsymbol{a}}(t) &= \boldsymbol{A} \boldsymbol{a}(t) + \boldsymbol{B} \boldsymbol{a}_{\text{in}}(t) \\
\boldsymbol{a}_{\text{out}}(t) &= \boldsymbol{C} \boldsymbol{a}(t) + \boldsymbol{D} \boldsymbol{a}_{\text{in}}(t)
\end{align}
where
\begin{widetext}
\begin{align}
  \boldsymbol{a} &= \left[\begin{matrix}{\hat a} & {\hat b} & {\hat c} &
      {{\hat a}^\dagger} & {{\hat b}^\dagger} & {{\hat c}^\dagger}\end{matrix}\right]^\top \\
  \boldsymbol{a}_{\text{in}} &= \left[\begin{matrix}{\hat a_{\text{in}}^c} & {\hat a_{\text{in}}^e} & {\hat b_{\text{in}}^c} & {\hat b_{\text{in}}^e} & {\hat c_{\text{in}}^e} & {(\hat a_{\text{in}}^c)^\dagger} & {(\hat a_{\text{in}}^e)^\dagger} & {(\hat b_{\text{in}}^c)^\dagger} & {(\hat b_{\text{in}}^e)^\dagger} & {(\hat c_{\text{in}}^e)^\dagger}\end{matrix}\right]^\top \\
\boldsymbol{a}_{\text{out}} &= \left[\begin{matrix}{\hat a_{\text{out}}^c} & {\hat b_{\text{out}}^c} & {(\hat a_{\text{out}}^c)^\dagger} & {(\hat b_{\text{out}}^c)^\dagger}\end{matrix}\right] \\
  \boldsymbol{A} &= \left[\begin{matrix}
      i \Delta_a - \frac{\kappa_{a}}{2} & 0 & i G_{a} & 0 & 0 & i G_{a}\\
      0 & i \Delta_b - \frac{\kappa_{b}}{2} & i G_{b} & 0 & 0 & i G_{b}\\
      i G_{a} & i G_{b} & - \frac{\gamma_{m}}{2} - i \omega_{m} & i G_{a} & i G_{b} & 0\\
      0 & 0 & - i G_{a} & - i \Delta_a - \frac{\kappa_{a}}{2} & 0 & - i G_{a} \\
      0 & 0 & - i G_{b} & 0 & - i \Delta_b - \frac{\kappa_{b}}{2} & - i G_{b} \\
      - i G_{a} & - i G_{b} & 0 & - i G_{a} & - i G_{b} & - \frac{\gamma_{m}}{2} + i \omega_{m}
    \end{matrix}\right] \\
  \boldsymbol{B} &= \left[\begin{matrix}
      \sqrt{\kappa_{a}^c} & \sqrt{\kappa_{a}^e} & 0 & 0 & 0 & 0 & 0 & 0 & 0 & 0\\
      0 & 0 & \sqrt{\kappa_{b}^c} & \sqrt{\kappa_{}^e} & 0 & 0 & 0 & 0 & 0 & 0\\
      0 & 0 & 0 & 0 & \sqrt{\gamma_{m}} & 0 & 0 & 0 & 0 & 0\\
      0 & 0 & 0 & 0 & 0 & \sqrt{\kappa_{a}^c} & \sqrt{\kappa_{a}^e} & 0 & 0 & 0\\
      0 & 0 & 0 & 0 & 0 & 0 & 0 & \sqrt{\kappa_{b}^c} & \sqrt{\kappa_{b}^e} & 0\\
      0 & 0 & 0 & 0 & 0 & 0 & 0 & 0 & 0 & \sqrt{\gamma_{m}}
    \end{matrix}\right] \\
  \boldsymbol{C} &= \left[\begin{matrix}
      \sqrt{\kappa_{a}^c} & 0 & 0 & 0 & 0 & 0\\
      0 & \sqrt{\kappa_{b}^c} & 0 & 0 & 0 & 0\\
      0 & 0 & 0 & \sqrt{\kappa_{a}}^c & 0 & 0\\
      0 & 0 & 0 & 0 & \sqrt{\kappa_{b}^c} & 0
    \end{matrix}\right] \\
  \boldsymbol{D} &= \left[\begin{matrix}-1 & 0 & 0 & 0 & 0 & 0 & 0 & 0 & 0 & 0\\0 & 0 & -1 & 0 & 0 & 0 & 0 & 0 & 0 & 0\\0 & 0 & 0 & 0 & 0 & -1 & 0 & 0 & 0 & 0\\0 & 0 & 0 & 0 & 0 & 0 & 0 & -1 & 0 & 0\end{matrix}\right]
\end{align}
\end{widetext}

This state space model can be transformed to the frequency-space transfer function $\boldsymbol{\Xi}(\omega)$ relating the input frequency modes $\boldsymbol{a}_{\text{in}}(\omega)$ to the output frequency modes $\boldsymbol{a}_{\text{out}}(\omega)$ in the steady-state
\begin{align}
\boldsymbol{a}_{\text{out}}(\omega) &= \boldsymbol{\Xi}(\omega) \boldsymbol{a}_{\text{in}}(\omega) \\
\boldsymbol{\Xi}(\omega) &= \boldsymbol{C}(-i\omega \boldsymbol{I}_6 - \boldsymbol{A})^{-1} \boldsymbol{B} + \boldsymbol{D}
\end{align}
where $\boldsymbol{I}_6$ is the $6\times 6$ identity matrix.

In principle $\boldsymbol{\Xi}(\omega)$ describes the unitary evolution of the system and environmental modes, however as the environmental modes will be traced over, their output modes are neglected in $\boldsymbol{a}_{\text{out}}$, simplifying $\boldsymbol{\Xi}(\omega)$ to a rectangular $4\times 10$ complex matrix.
In order to make the rest of the analysis symbolically tractable, we apply the approximations discussed in section \ref{sec:deviceModel} to get $\boldsymbol{\Xi}(\omega=\pm\omega_m)$.
As $\boldsymbol{\Xi}(\omega=\pm\omega_m)$ preserves the canonical commutation relations of the system and environmental operators, it is a linear symplectic map, but in complex form \cite{weedbrook12_gauss_quant_infor, adesso14_contin_variab_quant_infor}.
For convenience, we transform $\boldsymbol{\Xi}(\omega=\pm\omega_m)$ to the more standard quadrature basis symplectic form to get  $\boldsymbol{\Xi}^{xp}(\omega=\pm\omega_m)$ which is now defined by the vector of quadature operators given by
\begin{widetext}
\begin{align}
  \boldsymbol{a}_{\text{in}}^{xp} &= \left[\begin{matrix}
      {\hat x_{a,\text{in}}^c} & {\hat p_{a,\text{in}}^c} & {\hat x_{b,\text{in}}^c} & {\hat p_{b,\text{in}}^c} & {\hat x_{a,\text{in}}^e} & {\hat p_{a,\text{in}}^e} & {\hat x_{b,\text{in}}^e} & {\hat p_{b,\text{in}}^e} & {\hat x_{c,\text{in}}^e} & {\hat p_{c,\text{in}}^e}
      \end{matrix}\right]^\top \\
  \boldsymbol{a}_{\text{out}}^{xp} &= \left[\begin{matrix}
      {\hat x_{a,\text{out}}^c} & {\hat p_{a,\text{out}}^c} & {\hat x_{b,\text{out}}^c} & {\hat p_{b,\text{out}}^c}
    \end{matrix}\right]^\top .
\end{align}
\end{widetext}

We then reduce $\boldsymbol{\Xi}^{xp}(\omega=\pm\omega_m)$ to two-mode Gaussian channels acting only on the system OE modes which are described by the unitary-like $4\times 4$ matrix $\boldsymbol{T}$ and the $4\times 4$ noise-like matrix $\boldsymbol{N}$. We rearranged the ordering of the operators in going from $\boldsymbol{a}_{\text{in}}$ to $\boldsymbol{a}_{\text{in}}^{xp}$ so now we have that $\boldsymbol{T}$ is simply given by the $4\times 4$ matrix of the first four columns of $\boldsymbol{\Xi}^{xp}(\omega=\pm\omega_m)$ defined by the system input operators.
Then, $\boldsymbol{N}$ is found by first taking the $4\times 6$ matrix, which we will call $\boldsymbol{M}$, of the remaining six columns of $\boldsymbol{\Xi}^{xp}(\omega=\pm\omega_m)$ defined by the environmental bath operators and then computing $\boldsymbol{N} = \boldsymbol{M\Sigma M}^\top$ where $\boldsymbol{\Sigma} = \boldsymbol{I}_6/2 + n_{\text{th}}(\boldsymbol{0}_4 \oplus \boldsymbol{I}_2)$.

The full form of $\boldsymbol{T}$ and $\boldsymbol{N}$ is finally given by
\begin{widetext}
\begin{align*}
    \boldsymbol{T}_{\sigma_a,\sigma_b}
    &=
    \frac{1}{1-\sigma_a C_a-\sigma_b C_b}
    \\ &\qquad \times
    \begin{pmatrix}
        \sqrt{\delta_a \epsilon_a}(-1+\sigma_a C_a+\sigma_b C_b + 2\tau_a(1-\sigma_b C_b))\boldsymbol{I}_2
        &
        2\sqrt{\epsilon_a \delta_b \tau_a \tau_b C_a C_b} \begin{pmatrix}\sigma_a & 0 \\ 0 & \sigma_b\end{pmatrix}
        \\
        2\sqrt{\delta_a \epsilon_b \tau_a \tau_b C_a C_b} \begin{pmatrix}\sigma_b & 0 \\ 0 & \sigma_a\end{pmatrix}
        &
        \sqrt{\delta_b \epsilon_b} (-1 + \sigma_a C_a + \sigma_b C_b + 2\tau_b(1-\sigma_a C_a))\boldsymbol{I}_2
    \end{pmatrix}
    \\
    \boldsymbol{N}_{\sigma_a,\sigma_b}
    &=
    \frac{1}{2}
    \begin{pmatrix}
        \mu \boldsymbol{I}_2
        &
        \gamma \begin{pmatrix}\sigma_a \sigma_b & 0 \\ 0 & 1 \end{pmatrix}
        \\
        \gamma \begin{pmatrix}\sigma_a\sigma_b & 0 \\ 0 & 1\end{pmatrix}
        &
        \nu \boldsymbol{I}_2
    \end{pmatrix}
    \\
    &\qquad \mu = 1-\frac{\epsilon_a\left( \delta_a (1 - \sigma_a C_a - \sigma_b C_b - 2\tau_a (1-\sigma_b C_b))^2 - 4\tau_a C_a (1+\sigma_a+2n_\text{th}+C_b(2+\sigma_a+\sigma_b-\delta_b \tau_b)) \right)}{(1 - \sigma_a Ca - \sigma_b C_b)^2}
    \\
    &\qquad \nu = 1-\frac{\epsilon_b\left( \delta_b (1 - \sigma_a C_a - \sigma_b C_b - 2\tau_b (1-\sigma_a C_a))^2 - 4\tau_b C_b (1+\sigma_b+2n_\text{th}+C_a(2+\sigma_a+\sigma_b-\delta_a \tau_a)) \right)}{(1 - \sigma_a Ca - \sigma_b C_b)^2}
    \\
    &\qquad \gamma = \frac{2\sqrt{\epsilon_a \epsilon_b \tau_a \tau_b C_a C_b}}{(1 - \sigma_a Ca - \sigma_b C_b)^2} 
                \left[
                    4n_\text{th} + (\sigma_a \delta_a + \sigma_b \delta_b) (1 - \sigma_a Ca - \sigma_b C_b) + (1-\sigma_a \sigma_b)(1+C_a+C_b)
                \right. \\ &\qquad\qquad \left.
                    - 2\sigma_a \delta_a \tau_a (1-\sigma_b C_b) 
                    - 2\sigma_b \delta_b \tau_b (1-\sigma_a C_a) 
                \right]
\end{align*}
\end{widetext}
where $\sigma_a$ and $\sigma_b$ are the sign of the optical and microwave pump detunings respectively ($-1$ for red detuned, $+1$ for blue detuned).

\section{Derivation of the PPT criteria for the Two-Mode Beamsplitter-Type Interaction \label{sec:ppt-threshold}}

Here we find the constraints on transducer parameters in order for a single DPT operated with the beamsplitter-type interaction (i.e. both pumps red detuned) to facilitate the production of distillable entanglement (i.e. NPT states \cite{giedke01_entan_criter_all_bipar_gauss_states}) shared between optical and microwave frequency bosons.
Let $\mathcal{E}$ be the 2-mode channel implemented by the beamsplitter-type interaction of a DPT (this channel is completely characterized by $\boldsymbol{T}_{-1,-1}$ and $\boldsymbol{N}_{-1, -1}$).
Next, the Choi state $\hat\rho$ is found by initializing both the $A_1$-$A_2$ and the $B_1$-$B_2$ systems in infinitely squeezed two-mode squeezed vacuum states, and then having the channel $\mathcal{E}$ act on the $A_1$ and $B_1$ modes (see Fig. \ref{fig:two-mode}).
As a consequence of the Choi–Jamiołkowski (CJ) isomorphism, the channel $\mathcal{E}$ can only output states which are PPT (i.e., cannot output NPT states) if and only if the Choi state $\hat{\rho}$ is PPT (see Appendix \ref{sec:ppt-proof}).
Therefore we will find the conditions under which $\hat{\rho}$ is PPT.
We start by constructing the covariance matrix $\boldsymbol{V}$ of the state $\hat{\rho}$ which is given by
\begin{align}
    \boldsymbol{V} &= \boldsymbol{T'} \boldsymbol{V}_{in} \boldsymbol{T'}^\top + \boldsymbol{N'}
\end{align}
where from Fig. \ref{fig:two-mode} we see that
\begin{align}
    \boldsymbol{T'} &= \boldsymbol{I}_2 \oplus \boldsymbol{T} \oplus \boldsymbol{I}_2
    \\
    \boldsymbol{N'} &= \boldsymbol{0}_2 \oplus \boldsymbol{N} \oplus \boldsymbol{0}_2
    \\
    \boldsymbol{V}_{in} &= \frac{1}{2} \bigoplus_{j=1,2} \begin{pmatrix} \boldsymbol{I}_2\cosh r & \boldsymbol{Z}_2\sinh r \\ \boldsymbol{Z}_2\sinh r & \boldsymbol{I}_2 \cosh r\end{pmatrix}
\end{align}
and $\boldsymbol{T} = \boldsymbol{T}_{-1,-1}$ and $\boldsymbol{N} = \boldsymbol{N}_{-1,-1}$ since we are working with a beamsplitter-type transducer.
Additionally, $\boldsymbol{I}_2$ and $\boldsymbol{0}_2$ are the $2\times 2$ identity and zero matrix respectively, and $\boldsymbol{Z}_2=\text{diag}(1,-1)$.
The CJ Isomorphism requires taking $r\to\infty$; however, as we will see later, this will not in fact be necessary to do here here.

The state $\hat{\rho}$ is PPT if and only if \cite{weedbrook12_gauss_quant_infor}
\begin{align}
    \label{eqn:general-PPT-condition}
    \boldsymbol{\widetilde{V}} + i\boldsymbol{\Omega} \geq 0   
\end{align}
where
\begin{align}
    \boldsymbol{\Omega} = \frac{1}{2} \bigoplus_{j=1}^4 \begin{pmatrix} 0 & 1 \\ -1 & 0 \end{pmatrix}
\end{align}
and $\boldsymbol{\widetilde{V}}$ is the partial transpose of $\boldsymbol{V}$ with respect to the optical modes $A_1,A_2$ which is given by
\begin{align}
    \boldsymbol{\widetilde{V}} &= \boldsymbol{P} \boldsymbol{V} \boldsymbol{P}^\top
\end{align}
where $\boldsymbol{P} = \boldsymbol{Z}_2 \oplus \boldsymbol{Z}_2 \oplus \boldsymbol{I}_4$, and $\boldsymbol{I}_4$ is the $4\times 4$ identity matrix.

Nominally, conditions under which $\hat{\rho}$ is PPT can be found by using Eq.~\eqref{eqn:general-PPT-condition}.  This would be done by finding the eigenvalues $\{\lambda_i\}$ of the matrix $\boldsymbol{\widetilde{V}} + i\boldsymbol{\Omega}$ and requiring $\lambda_i \geq 0$ for all $i$.  Stated explicitly
\begin{align}
    \left\{ \lambda_i \geq 0 \left| \det \left( \boldsymbol{\widetilde{V}} + i\boldsymbol{\Omega} - \lambda_i \boldsymbol{I}_8 \right) = 0 \right. \right\}
\end{align}
However, since the eigenvalues are the zeros of the 8th order characteristic polynomial, this method returns unwieldy expressions for $\lambda_i$ which are difficult to simplify.
An alternative method, which is computationally simpler, is to first find the boundary between PPT and NPT regions and then to identify the PPT and NPT sides.

We start by finding the conditions on the physical parameters which cause one or more of the eigenvalues to be zero.
This is true if and only if the determinant is zero, so the boundary is defined by the equation
\begin{align}
    \det \left( \boldsymbol{\widetilde{V}} + i\boldsymbol{\Omega} \right) = 0
\end{align}
Upon solving this equation for $n_{\text{th}}$ we find there are two solutions
\begin{align}
    n_\text{th} \in \left\{ \nu_a, \nu_b \right\}
\end{align}
where
\begin{align}
  \nu_i = \frac{\delta_i \tau_i C_i}{1 - \delta_i\epsilon_i(1 - 2\tau_i)^2}
\end{align}
We find that when $\min \left\{ \nu_a, \nu_b \right\} < n_{\text{th}} < \max \left\{ \nu_a, \nu_b \right\}$ one eigenvalue is negative, whereas when $n_{\text{th}} < \min \left\{ \nu_a, \nu_b \right\}$ two eigenvalues are negative, and when $n_{\text{th}} > \max \left\{ \nu_a, \nu_b \right\}$ no eigenvalue is negative.
Therefore both $\hat{\rho}$ and $\mathcal{E}$ are PPT if and only if
\begin{align}
  n_{\text{th}} \ge \max \left\{ \nu_a, \nu_b \right\} .
\end{align}
Note that we did not need to take the limit $r\to\infty$ because the threshold is independent of $r$ (provided $r\neq 0$).

\section{A Channel is PPT-Preserving if and only if its Choi State is PPT \label{sec:ppt-proof}}

In this section we prove a slight generalization of proposition (iii) in \cite{cirac01_entan_operat_their_implem_using}.
Let us consider a completely positive map (i.e., channel) $\mathcal{E}$ acting on systems $A_1$ and $B_1$, while implicitly understanding it to act as the identity on all other modes.
Let $E_{A_1 A_2, B_1 B_2}$, acting on $\mathcal{H}_{A_1} \otimes \mathcal{H}_{A_2} \otimes \mathcal{H}_{B_1} \otimes \mathcal{H}_{B_2}$ (where $\dim(H_{A_i}) = \dim(H_{B_i}) = d$), be the Choi state of $\mathcal{E}$ which is defined as \cite{cirac01_entan_operat_their_implem_using, giedke02_charac_gauss_operat_distil_gauss_states}
\begin{equation}
E_{A_1 A_2, B_1 B_2} = \mathcal{E}\left(\ket{\Psi}\bra{\Psi}_{A_1 A_2} \otimes \ket{\Psi}\bra{\Psi}_{B_1 B_2}\right)
\end{equation}
where $\ket{\Psi}$ is the maximally entangled state defined as
\begin{equation}
\ket{\Psi}_{X Y} = \frac{1}{\sqrt{d}} \sum_{i=1}^d \ket{i}_X \otimes \ket{i}_Y
\end{equation}
and $\{\ket{i}\}_{i=1}^d$ is an orthonormal basis.
Finally we define the additional systems $A'$ and $B'$ with $\dim(\mathcal{H}_{A'_i}) = \dim(\mathcal{H}_{B'_i}) = d'$ so that we can write a density operator $\rho_{A_1 A' B_1 B'}$ acting on $\mathcal{H}_{A_1} \otimes \mathcal{H}_{A'} \otimes \mathcal{H}_{B_1} \otimes \mathcal{H}_{B'}$ that will have modes the channel $\mathcal{E}$ acts on while containing arbitrary modes in systems $A'$ and $B'$ that $\mathcal{E}$ acts as identity on.
We also define systems $A''$ and $B''$ similarly to $A'$ and $B'$.

Now we can write down the statement which we will prove:
\begin{quote}
  For all $\rho_{A_1 A' B_1 B'}^{\top_{A_1 A'}} \ge 0$, we have that $\mathcal{E}(\rho_{A_1 A' B_1 B'})^{\top_{A_1 A'}} \ge 0$ if and only if $E_{A_1 A_2, B_1 B_2}^{\top_{A_1 A_2}} \ge 0$.
\end{quote}
where $\top_{X}$ refers to partial transposition with respect to subsystem $X$.

The forward implication is trivial to show since we can write $\rho_{A_1 A' B_1 B'} = \ket{\Psi}\bra{\Psi}_{A_1 A_2} \otimes \ket{\Psi}\bra{\Psi}_{B_1 B_2}$ where we let $A_2 B_2$ and $A' B'$ refer to the same ancillary system.

In order to prove the reverse implication we will need the following relation which one can readily show. 
\begin{widetext}
\begin{align}
  \mathcal{E}(\rho_{A_1 A' B_1 B'})
  &= d^2 d'^2 \tr_{A_2 A'' B_2 B''}\big(E_{A_1 A_2, B_1 B_2} 
    \ket{\Psi}\bra{\Psi}_{A' A''} \ket{\Psi}\bra{\Psi}_{B' B''}
    \rho_{A_2 A'', B_2 B''}^{\top_{A_2 A'' B_2 B''}}\big)
\end{align}
Immediately following from the above relation we have
\begin{align}
  \mathcal{E}(\rho_{A_1 A' B_1 B'})^{\top_{A_1 A'}}
  &= d^2 d'^2 \tr_{A_2 A'' B_2 B''}\left(E_{A_1 A_2, B_1 B_2}^{\top_{A_1 A_2}}
    \ket{\Psi}\bra{\Psi}_{A' A''} \ket{\Psi}\bra{\Psi}_{B' B''}
    \rho_{A_2 A'', B_2 B''}^{\top_{B_2 B''}}\right)
\label{eqn:cirac01-eqn8-generalized}
\end{align}
Now if we consider an arbitrary $\ket{\psi}_{A_1 A', B_1 B'} \in \mathcal{H}_{A_1 A', B_1 B'}$ we get that
\begin{align}
  \bra{\psi}_{A_1 A', B_1 B'} &\mathcal{E}(\rho_{A_1 A' B_1 B'})^{\top_{A_1 A'}} \ket{\psi}_{A_1 A', B_1 B'}
  \nonumber \\&=
  d^2 d'^2 \tr\left(\left(E_{A_1 A_2, B_1 B_2}^{\top_{A_1 A_2}}
    \ket{\Psi}\bra{\Psi}_{A' A''} \ket{\Psi}\bra{\Psi}_{B' B''} \right)
  \left(\rho_{A_2 A'', B_2 B''}^{\top_{B_2 B''}} \ket{\psi}\bra{\psi}_{A_1 A', B_1 B'} \right)\right) \label{eqn:final}
\end{align}
Recall that for any $\boldsymbol{M} \ge 0$ and $\boldsymbol{N} \ge 0$ we have $\boldsymbol{M}\otimes \boldsymbol{N} \ge 0$ and $\tr(\boldsymbol{M N}) \ge 0$. Therefore Eq.~\eqref{eqn:final} is non-negative, and so we conclude that $\mathcal{E}(\rho_{A_1 A' B_1 B'})^{T_{A_1 A'}} \ge 0$.

It's also worth stating the Choi–Jamiołkowski isomorphism in this more general context with additional, arbitrary ancilla modes
\begin{align}
  \mathcal{E}(\rho_{A_1 A' B_1 B'}) &=
  d^4 \tr_{A_2 A_3 B_2 B_3}\left(E_{A_1 A_2, B_1 B_2} \rho_{A_3 A' B_3 B'}
  \ket{\Psi}\bra{\Psi}_{A' A''} \ket{\Psi}\bra{\Psi}_{B' B''} \right) .
  \label{eqn:cirac01-eqn6}
\end{align}
The interpretation of this equation is the same as in \cite{cirac01_entan_operat_their_implem_using}, namely if we have the Choi state $E_{A_1 A_2, B_1 B_2}$, we can always (probabilistically) simulate the action of $\mathcal{E}$ on the $A_1 B_1$ subsystem $\rho_{A_1 A' B_1 B'}$ via the joint projection of $A_2 A_3$ subsystem onto the maximally entangled state while doing the same for the $B_2 B_3$ subsystem.

Note that our notion of PPT-preserving channels is equivalent to the definition of completely PPT-preserving channels in \cite{khatri20_princ_quant_commun_theor} where they also give a proof that a channel is completely PPT-preserving if and only if its Choi state is PPT with respect to the appropriate bipartition.
Proposition (iii) in \cite{cirac01_entan_operat_their_implem_using} lacked complete positivity which is necessary for the forward implication to hold, where a simple counterexample is that of the swap operator between systems $A$ and $B$.
Finally note that while this proof, along with those given in \cite{cirac01_entan_operat_their_implem_using, khatri20_princ_quant_commun_theor}, assumed finite dimensional Hilbert spaces of dimension $d$, it is expected to hold as $d\rightarrow\infty$ for Gaussian states and channels due to the CJ formalism for bosonic Gaussian systems established in \cite{holevo11_choi_jamiol_forms_quant_gauss_chann, holevo11_entrop_gain_choi_jamiol_corres}.
\end{widetext}

\bibliography{aiko-references}
\end{document}